\documentclass{sig-alternate-05-2015}
\usepackage{verbatim}
\usepackage{rotating}
\usepackage{xspace}
\usepackage{pifont}
\newcommand{\cmark}{\ding{51}}%
\newcommand{\xmark}{\ding{55}}%
\usepackage{pbox}
\usepackage{listings}
\usepackage{float}
\usepackage[font=small,labelfont=bf,textfont=bf]{caption}
\usepackage{multirow}
\usepackage{xcolor,colortbl}
\usepackage{tikz}

\newcommand*\circled[1]{\tikz[baseline=(char.base)]{
            \node[shape=circle,draw,color=white,fill=black,inner sep=1pt] (char){#1};}}

\newcolumntype{"}{@{\hskip\tabcolsep\vrule width 1pt\hskip\tabcolsep}}

\definecolor{Gray}{gray}{0.85}
\definecolor{Gray1}{gray}{0.65}
\definecolor{Gray2}{gray}{0.45}

\definecolor{LightCyan}{rgb}{0.88,1,1}

\newcommand{\system}{\textsf{AuDroid}\xspace}

\RequirePackage[normalem]{ulem} 
\RequirePackage{color}\definecolor{RED}{rgb}{1,0,0}\definecolor{BLUE}{rgb}{0,0,1} 

\begin{document}
%
\setcopyright{acmcopyright}
\conferenceinfo{ACSAC '15,}{December 07-11, 2015, Los Angeles, CA, USA}
\isbn{978-1-4503-3682-6/15/12}
\acmPrice{\$15.00}
\doi{http://dx.doi.org/10.1145/2818000.2818005}

\title{ {\ttlit AuDroid}: Preventing Attacks on Audio Channels \\ in Mobile Devices}

\numberofauthors{2} 
%


\author{
%
%
\alignauthor
Giuseppe Petracca, Yuqiong Sun, and Trent Jaeger \\ 
       \affaddr{Dept. of Computer Science and Engineering}\\
       \affaddr{The Pennsylvania State University}\\
       \affaddr{\{gxp18, yus138, tjaeger\}@cse.psu.edu}
\alignauthor
Ahmad Atamli \\ 
       \affaddr{Dept. of Computer Science}\\
       \affaddr{University of Oxford}\\
       \affaddr{atamli@cs.ox.ac.uk}
}


\maketitle

\begin{abstract}
Voice control is a popular way to operate mobile devices, enabling users to communicate requests to their devices. However, adversaries can leverage voice control to trick mobile devices into executing commands to leak secrets or to modify critical information.
Contemporary mobile operating systems fail to prevent such attacks because they do not control access to the speaker at all and fail to control {\em when} untrusted apps may use the microphone, enabling authorized apps to create exploitable communication channels.   
In this paper, we propose a security mechanism that tracks the creation of audio communication channels explicitly and controls the information flows over these channels to prevent several types of attacks.
We design and implement \system, an extension to the SELinux reference monitor integrated into the Android operating system for enforcing lattice security policies over the dynamically changing use of system audio resources.  To enhance flexibility, when information flow errors are detected, the device owner, system apps and services are given the opportunity to resolve information flow errors using known methods, enabling \system to run many configurations safely. We evaluate our approach on 17 widely-used apps that make extensive use of the microphone and speaker, finding that 
\system prevents six types of attack scenarios on audio channels while permitting all 17 apps to run effectively. 
\system shows that it is possible to prevent attacks using audio channels without compromising functionality or introducing significant performance overhead. 

\hfill
\\
\hfill
\end{abstract}

\vspace{-0.2in}


\section{Introduction}
Smartphones and wearable devices have changed the way people
interact with computers.  Where PCs require the keyboard and mouse to communicate, smartphones and wearables enable interaction using other inputs, such as voice control.  For example, voice-activated ''personal assistants'' offer to help us with basic tasks in a hands-free way, including \emph{Siri} and \emph{Google Now}.  Other apps leverage the speaker and microphone on such devices for other purposes, such as voice messaging, phone call recording, and voice notes.

Unfortunately, adversaries may leverage access to the microphone or
speaker to launch a variety of attacks against apps and system services. Unauthorized voice recording has long been known to be a serious security issue.
For example, a malicious app may try to steal secret
information communicated to the smartphone \cite{talkback} (e.g., passwords entered by voice) or generated by the
smartphone \cite{texttospeech} (e.g., text converted to speech) by recording such information surreptitiously using the smartphone's microphone \cite{Diao}. 
In addition, a malicious app may replay recorded
or generated information using the device's speaker to gain access to
other secret information (e.g., through manufactured ''voice''
commands) or otherwise control the behavior of the device \cite{Diao,article}. 
Finally, even external adversaries, such as a user different from the device owner or in proximity mobile devices, may attack the
smartphone by submitting commands, tricking smartphone apps and services
into releasing secret information or modifying the device in an
unauthorized way.

Researchers have explored a variety of possible solutions to prevent
these attacks, but none of them is sufficient to
completely prevent exploitation of mobile devices. 
Research-ers have proposed
modifying individual apps to prevent attacks from untrusted
input to the microphone, such as the \emph{Google Voice Search} app~\cite{Diao},
but such techniques are app-specific and lack visibility into the system necessary to
determine whether an attack is possible.  \emph{SemaDroid}~\cite{Xu} enables
users to control the quality of sensor data collected and/or replace sensor data with mock data to prevent
privacy leaks, but such countermeasures may impact the effectiveness of apps.  
Researchers have also enhanced smartphones with access control systems
that enforce mandatory access control~\cite{seandroid,Smalley} (MAC),
reason about trusted paths with users~\cite{Jang,Roesner}, and enable
app-aware MAC~\cite{Heuser}, but these systems do not reason
about the communication channels created by individual accesses,
failing to prevent the attacks above.


The aim of this work is to control communication channels created when apps and services use the microphone and speaker to prevent
system apps (including services) from leaking secret information or
being used as a confused deputy ~\cite{deputy}. In theory, these attacks can be prevented by enforcing lattice security
policies over these channels~\cite{denning_lattice}.  By enforcing
simple two-level lattices for secrecy and integrity, we can prevent
data in system apps and services from being leaked to untrusted parties (e.g., third-party apps) and prevent
system apps and services from using data from untrusted parties, respectively.

However, there are two major challenges in enforcing such policies.
First, processes may take ownership of the microphone and/or speaker
dynamically, creating communication channels that may enable exploitation. 
We identify \emph{three distinct types of communication channels} that must be controlled:
(1) a channel from the device's speaker to its microphone; (2) a channel from the device's speaker to external parties who may be eavesdropping on the device; and (3) a channel from external parties, who may be trying to trick the device, to the device's microphone.  Any defense must track the creation and deletion of such channels and reason about the information flows created by such channels to prevent exploits.  
Second, enforcing a lattice policy may be too strict for some
apps, such as where users want to allow data to be captured by
third-party apps.  In such cases, we take two distinct approaches, depending on whether the channel may compromise the security of a system app/service or the device owner.  For system apps and services, we leverage available methods to resolve information flow errors in audio channels by declassifying or endorsing unsafe communications, which we collectively call (information flow) {\em resolvers}.  However, the effectiveness of resolvers may be application-specific, so we propose a method for negotiating the choice of resolvers using callbacks, motivated by the use of callbacks to enforce access control from Android Security Modules~\cite{Heuser}.  For users, we construct a trusted path to the Android operating system to notify the user of unsafe flows and obtain user approval of such flows, motivated by the creation of trusted paths for individual applications in
user-driven access control~\cite{Jang,Roesner}.  


In this paper, we design and implement \system, an extension to the SELinux reference monitor integrated into the Android OS to enforce lattice
policies over the dynamically changing use of system audio resources, and enlist input from system apps and services to evaluate options for resolving unsafe communication channels.
Specifically, we have integrated \system into the Android \emph{Media Server}
to enable it to control access to the microphone and speaker to
enforce a simple lattice policy that aims to protect system apps from
third-party apps and external attackers.  
We evaluate \system on \emph{six types of attack scenarios} described in section \ref{attacks}, including both previously published and new attacks.  We evaluate \system on 17 widely-used apps that leverage audio to verify that \system, when enhanced with available resolvers, prevents exploits without impairing the normal functionality of such system apps and services. 
We find that, as discussed in Section \ref{evaluation}, \system provides such defense for less than 4 $\mu$s for the speaker and less than 25$\mu$s for the microphone, even including visual notification, leading to insignificant overhead during app usage.

In summary, we make the following contributions:
\begin{itemize} 
\itemsep-0.3em 
  \item We classify \emph{six types of attack scenarios} that may leverage system audio resources to exploit system apps and services or the device owner, two of which are new types of attacks we identify in this paper.  
  \item To prevent these types of attacks, we propose \system, an extension to the SELinux reference monitor to enforce lattice security policies over three types of dynamically-created, audio communication channels.  \system adapts prior research that callbacks to apps and trusted paths to leverage available methods for resolving information flow errors, which we call {\em resolvers}, to permit restricted use of unsafe audio channels and prevent exploitation.
  \item We evaluate our approach on 17 widely-used apps that make extensive use of the microphone and speaker, finding that 
\system prevents all the discussed attack scenarios on audio channels while permitting all 17 applications to run effectively.
\end{itemize}

\vspace{-0.05in}

\section{Problem Definition}
Contemporary mobile operating systems do not provide mechanisms to manage audio streams securely. The main reason is because accesses to microphone and speaker, by different processes, are not considered security-sensitive. Currently, such operating systems only control access to the microphone, allowing authorized apps to access the microphone at any time, even when a privileged app may be using the speaker or the user may be speaking a sensitive passcode to the device.  Consequently, current mobile operating systems fail in enforcing information flow control through the microphone and speaker, resulting in putting the users' data confidentiality and device integrity at risk.  

In this paper, we focus on Android OS due to its open source availability and its wide adoption in the mobile device market \cite{market}. Without lack of generality, these observations could be extended to other mobile operating systems, currently adopted by millions of users worldwide, such as iOS and Windows Phone.

\subsection{Audio Stream Management in Android OS}
\label{android}

\begin{figure}[t]
\centering
\includegraphics[width=75mm]{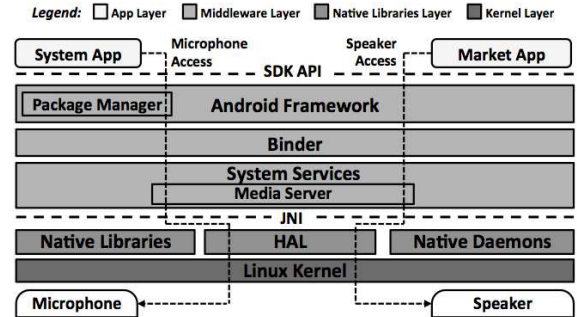}
\caption{Android OS Architecture}
\label{fig:arch}
\end{figure}

Figure~\ref{fig:arch} shows the architecture of the Android system and how Android apps request access to audio devices. Both Android \emph{system apps} (distributed with the OS) and \emph{market apps} (installed by the device owner) obtain access to Android system resources through requests via the SDK API to the \emph{system services} in the Android System Services layer.  In general, such system services use the Android Hardware Abstraction Layer (HAL) to obtain control of devices from the Linux kernel. For example, through the HAL library the Android \emph{Media Server} directly communicates with the underlying Linux kernel to read and write Linux device (i.e., \texttt{/dev}) files. SELinux rules on the Android system are configured to permit only the \emph{Media Server} to access the specific Linux device files for the speaker (\texttt{/dev/snd/pcmC0D15p}) and the microphone (\texttt{/dev/snd/pcmC0D0c}). Therefore, the process running the \emph{Media Server} is the only one allowed to directly access the microphone and speaker.

Currently, the Android system only controls apps' access to the microphone device, not the speaker.  The \emph{Media Server} uses the Android permission mechanism to authorize access to system resources. To access the microphone, apps need to declare the following security permission, \texttt{<uses-permission android:name="android.permission.RECORD\_ AUDIO"/>}, in their \texttt{AndroidManifest.xml} file \cite{permissions}. This permission is validated by the \emph{Package Manager}, a module of the Android framework, both at app installation time and each time the app requests access to the microphone.  
If an app attempts to access the microphone without specifying the permission in the manifest file, the \emph{Package Manager} throws a security exception and terminates the app execution. On the other hand, to access the speaker no specific security permission needs to be declared by apps, so any app can use the speaker any time it is available. 

\subsection{Audio Channels and Attack Scenarios} \label{attacks}

{\em Audio channels} enable one principal who is sending audio signals to communicate with another principal who receives these signals.  Normally, apps and services use audio channels to communicate with external parties.  For example, apps and services may use the microphone to receive commands from the device owner and may use the speaker to play music or produce sounds directed to the device owner.  Given that smartphones have a microphone and a speaker and communicate with external parties, we identify the following \emph{three types of audio channels} that may be created: 

\begin{itemize}
\itemsep0em
\item \textbf{Channel Type 1} - from the device's speaker to the device's microphone. 
\item \textbf{Channel Type 2} - from the device's speaker to an external party (e.g., user or other device). 
\item \textbf{Channel Type 3} - from an external party (e.g., user or other device) to the device's microphone.  
\end{itemize}

\begin{figure}
\centering
\includegraphics[width=75mm]{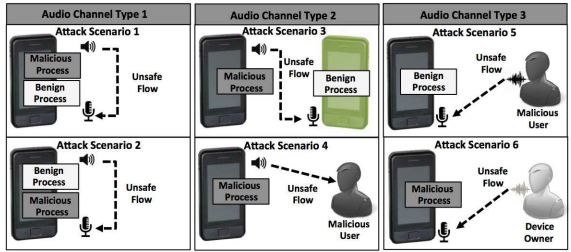}
\caption{Six different types of attack scenarios}
\label{fig:channels}
\end{figure}

Unfortunately, audio channels may also be misused to compromise the secrecy of the device owner or trick the device into performing unauthorized operations (i.e., use the \emph{Media Server} as a {\em confused deputy}~\cite{deputy}).  Figure~\ref{fig:channels} shows six different ways that audio channels may be exploited, as detailed below\footnote{Scenarios where other devices are the external parties can be reduced to one of the six attack scenarios identified.}.

\begin{itemize}
\itemsep-0.3em

\item \textbf{Attack Scenario 1}. A malicious process uses the device's speaker to produce input, such as malicious voice commands, to trick a benign (victim) process using the device's microphone into acting as a confused deputy \cite{deputy}. 
\item \textbf{Attack Scenario 2}. A malicious process uses the device's microphone to eavesdrop on the output from the device's speaker, produced by a benign process, and record security-sensitive information.
\item \textbf{Attack Scenario 3}. A malicious process uses the device's speaker to send input to external parties, such as another mobile device or the user, to trick them into acting as a confused deputy.
\item \textbf{Attack Scenario 4}. In this scenario both ends are malicious. A malicious process sends secret information from the device's speaker that may be received by a malicious external party listening to the device. 
\item \textbf{Attack Scenario 5}. A malicious external party produces input, such as malicious voice commands, to trick a benign  process using the device's microphone into acting as a confused deputy. 
\item \textbf{Attack Scenario 6}. A malicious process uses the device's microphone to eavesdrop on external parties, such as the device owner.  
\end{itemize}
These six types of attack scenarios represent one eavesdropping and one confused deputy attack for each of the three types of audio channels.  Researchers have previously identified how some of these attacks may be realized.  Diao \emph{et al.} \cite{Diao} demonstrate how a malicious app bypasses the \emph{Touchless Voice Authentication}, a mechanism that authenticates the user, to unlock the screen and start the \emph{Google Voice Search} (GVS) app by using voice commands (Scenario 1).  
Jang \emph{et al.} \cite{Jang} show how a keylogger installed on the target device eavesdrops on the keyboard inputs and steal secrets, such as passwords, when typed by the device owner and spoken out by the \emph{TalkBack} assistive technology \cite{talkback} (Scenario 2). 
Furthermore, a malicious app could use the microphone to eavesdrop on the external party to collect security-sensitive information, such as private conversations and phone calls.   Schlegel \emph{et al.} \cite{Schlegel} show that apps  with  the \texttt{RECORD\_AUDIO} permission can selectively extract confidential  data  (e.g., credit  card  numbers) and  stealthily  deliver  them  to  the  adversary  (Scenario 6).

\vspace{+0.2in}

\begin{table}[H]
\centering
\caption{Security-sensitive Voice Command Examples} 
\label{table:commands}
\tiny
\begin{tabular}{|c|l|c|}
\hline
\multicolumn{1}{|c|}{\textbf{Device}} & \multicolumn{1}{|l|}{\textbf{Voice Command}}  \\ \hline

\multirow{6}{*}{\begin{turn}{90} Android OS \end{turn}}                                  
& When's my next meeting? \\ \cline{2-2} 
& Video call Jane using Hangouts \\ \cline{2-2}                                         
& Listen to voicemail \\ \cline{2-2} 
& Activate \textit{Speak Passwords} in accessibility settings \\ \cline{2-2} 
& Show me my flights \\ \cline{2-2} 
& Send an email to Kristin, subject ..., message, ...  \\ \hline

\multirow{6}{*}{\begin{turn}{90} iOS \end{turn}}                                         
& Open my Facebook \\ \cline{2-2} 
& Call 800 600 1234 \\ \cline{2-2}                                                      
& What's going on Twitter? \\ \cline{2-2} 
& Read my last email \\ \cline{2-2} 
& Show me my pictures \\ \cline{2-2} 
& Cancel my morning alarm for tomorrow  \\ \hline

\multirow{4}{*}{\begin{turn}{90} \begin{tabular}[x]{@{}c@{}}Amazon\\Echo\end{tabular}  \end{turn}} 
& Alexa, buy this song \\ \cline{2-2} 
& Alexa, what's my commute? \\ \cline{2-2}
& Alexa, turn off Bedroom Switch. \\ \cline{2-2} 
& Alexa, send this Voicecast on Sarah's tablet \\ \hline
\end{tabular}
\end{table}


In addition, we show that the other three types of attack scenarios listed are also possible. For example, as described in Section \ref{evaluation}, we developed a malicious app that plays voice commands (such as those listed in Table \ref{table:commands}) to external parties, specifically other devices, such as other smartphones, tablets, thermostats, home security systems, or voice-controlled automobile systems (Scenario 3). We also developed an app that stealthy eavesdrops voice and sound through the microphone to collect security-sensitive information such as private  conversations,  successively  leaked  to  an  adversary through  the  speaker  as  soon  as  the  device  owner,  victim of  the  attack,  is  distant  from  the  device (Scenario 4). This is essentially a Trojan horse attack from the device. Finally, attacks due to a misplaced trust in external party's input are possible whenever a malicious user is able to send malicious voice commands to system apps or services (Scenario 5).

\vspace{-0.05in}

\subsection{Challenges in Protecting Audio Channels}

While researchers have identified the possibility of such attacks, systematic defenses to prevent such attacks have not been developed yet.  We identify three challenges in protecting communications via audio channels:

\begin{itemize}
\itemsep-0.3em 




\item Mobile devices enable dynamic creation of audio channels that might not comply with information flow requirements, enabling both eavesdropping and confused deputy attacks. 


\item The functional requirements of some apps may violate information flow requirements, causing errors when traditional information flow control defenses are used.  

\item Audio channels may enable communication with external parties, such as other devices and users, whose identity and intentions may not be known.


\end{itemize}





Proposed security mechanisms do not address these challenges.  First, several proposed security mechanisms are app-specific.  Researchers propose removing vulnerabilities from system apps and services, such as Google Voice Search (GVS) \cite{Diao}, but there are several such system apps and services.  Researchers have proposed blocking access to the speaker while a system service is actively listening for voice commands \cite{Diao, Jang}, but this solution may impact usability (prevent playing of ring tones or notification sounds) and does not address control of external parties. 
Researchers have proposed solutions to protect secrecy of sensor information by returning fake or fuzzed sensor data to apps~\cite{Xu}.  While this approach works reasonably for some sensors, such as GPS and the accelerometer, it does not suffice for audio data as apps require accurate audio data to perform their tasks.
The Android permission mechanism~\cite{permissions} and Security Enhanced Linux for Android (SE Android) \cite{seandroid,Smalley} cannot prevent such attacks.  As discussed above, neither controls access to the speaker at present. In addition, neither considers audio channels as objects in need of control.  As a result, a malicious app that has been granted access to the microphone may use that privilege whenever the microphone is available. 

An interesting research proposal is to leverage app-specific information to augment access control.  For example, researchers have proposed using callbacks to app-trusted modules to evaluate unsafe operations~\cite{Heuser,Backes} and using trusted modules to track user interaction with the device~\cite{Roesner}.  For example, the Android Security Modules (ASM) framework~\cite{Heuser} allows apps to register for a callback for specific authorization hooks.  However, only the apps using audio channels should be notified and such callbacks should also enable the resolution of information flow errors in addition to authorization decisions.   
Also, User-Driven Access Control (UDAC) \cite{Roesner} enables untrusted apps to choose when to use trusted modules for the creation of a trusted path.  Unfortunately, that requires that the reference monitor to rely on the app developers to create the trusted paths or risk breaking functionality.  However, we find both the ASM and UDAC ideas interesting, and propose enhancements that enable effective control of audio channels.


\vspace{-0.05in}

\section{Security Model} \label{model}
\subsection{Threat Model}

In our threat model we consider two sources of threats. The first source of threats is {\em internal} to the mobile device: processes running market apps installed by the device owner.  We assume that the device owner is not aware of the maliciousness of market apps.  The second source of threats is {\em external} to the device: external parties, including users other than the device owner and other devices. 
Malicious apps, users, and devices can perform any of the attacks described in Section \ref{attacks}: eavesdropping and confused deputy attacks on the three types of audio channels.

\subsection{Trust Model}

We assume that the operating system of the target device (e.g., Linux kernel and Android OS) is booted securely (e.g., \emph{Verified Boot} mechanism \cite{boot}), runs approved code from device vendor, is free of malice, and is trusted to protect itself from the threats above.  We assume that system apps and services run approved code from the device vendor and are free of malice.  Some market apps might contain native code that may try to gain access to system/hardware resources. To prevent such native code from being able to directly access system/hardware resources, we rely on the use of \emph{SELinux} \cite{seandroid} running in \emph{Enforcing Mode} from boot time.  Therefore, only system services can access physical devices through the use of the Java Native Interface (JNI) \cite{jni}.  We assume that SELinux for Android satisfies the reference monitor concept~\cite{monitor}.  Thus, we assume market apps can only access the microphone and the speaker through the API provided by the standard Android SDK \cite{sdk}.  
\vspace{-0.05in}

\section{{\secit \large \system} design} 
\label{sect:design}
\begin{figure} [t]
\centering
\includegraphics[width=80mm]{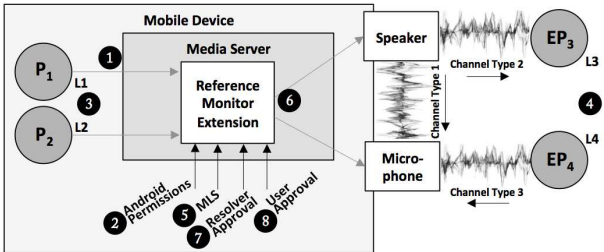}
\caption{Overview of the \system Design}
\label{fig:overview}
\end{figure}

In this section, we detail the design of \system, an extension to the SELinux
reference monitor, for enforcing Multi-Level Security (MLS)
over audio channels created by apps and system services, while using the microphone and speaker, to prevent the attacks described in Section~\ref{attacks}.  One key insight of designing  \system is to authorize access to all three types of dynamically-created audio channels highlighted in Figure~\ref{fig:overview}.  \system extends the reference monitor provided by SELinux by placing 4 additional hooks in the \emph{Media Server} to mediate all accesses to the microphone and speaker by apps and services, compute the audio channels that would be created by the access, and authorize the resultant information flows of those audio channels if they comply with an MLS policy. A second insight of designing \system is to create {\em trusted paths} with the device owners, authenticate them and learn their intentions when attempting to use unsafe information flows.  Additionally, to improve flexibility, \system  enables the use of available methods to resolve some information flow errors, which we will call (information flow) {\em resolvers}. 


\subsection{{\secit \system} Overview}
\label{overview}

An overview of the \system design is shown in Figure~\ref{fig:overview}. \system implements an extension to the reference monitor provided by SELinux to mediate accesses to the microphone and speaker by enhancing the {\em Media Server}.  The reference monitor extension is activated upon receiving a request from a process, i.e., one of the two internal parties ($P_{1}$ and $P_{2}$), to access either the microphone or the speaker (step~\circled{1}). If the access request is for the microphone, the reference monitor first checks the Android permissions (step~\circled{2}), as it does already. Then, the {\em Media Server} reasons about which of the three possible audio channels would be created by considering the possible external parties ($EP_{3}$ and $EP_{4}$), and identifies the security levels ($L1$, $L2$, $L3$, and $L4$) associated with the parties that would communicate as result of granting access to the microphone or speaker (steps~\circled{3} and \circled{4}), as described in Section~\ref{levels}. 
Subsequently, the {\em Media Server} enforces an MLS policy over the three audio channels (step~\circled{5}), as described in Section~\ref{mls}.  If, for all three audio channels, the corresponding information flows are classified as safe by the MLS policy, then the access request is granted and the corresponding audio channels are created (step~\circled{6}).  Otherwise, if an information flow is identified as unsafe, \system has two ways to determine whether the unsafe information can be resolved (i.e., made sufficiently safe to prevent eavesdropping and/or confused deputy attacks).  First, \system proposes a known {\em resolver} and requires approval from the system app or service at risk to verify that the proposed resolver will protect the system app or service and enable it to function (step~\circled{7}), as described in Section~\ref{resolver}.  In addition, if the unsafe information flows involve the use of the microphone, \system creates a {\em trusted path} for the device owner to confirm the access to the microphone is acceptable (e.g., recording is not eavesdropping and/or output is understood to be from a low-integrity source) to the device owner (step~\circled{8}), as described in Section~\ref{trusted}. \system gathers system apps/services and the device owner willingness to be part of the communication channels and accordingly modifies the security levels to resolve information flow errors. Neither the user nor the system apps/services are allowed to change the policy under enforcement.

\subsection{Identification of Security Levels} \label{levels}

In order to mediate accesses to the microphone and speaker, it is necessary to identify the security levels associated to the parties involved in all three audio channels. 
\system identifies the security levels of each internal party by using the process ID (PID) specified in the access requests for the microphone or speaker. \system leverages the convention used by the Linux kernel in Android OS, according to which market apps have PID greater than 2001, system apps have PID between 1001 and 2000, and system services have PID between 1 and 1000. According to this convention, \system identifies system apps and services as high-secrecy and high-integrity \texttt{(HS,HI)} subjects, market apps are identified as low-secrecy and low-integrity \texttt{(LS,LI)} subjects.

\begin{figure}[h]
\centering
\includegraphics[width=65mm]{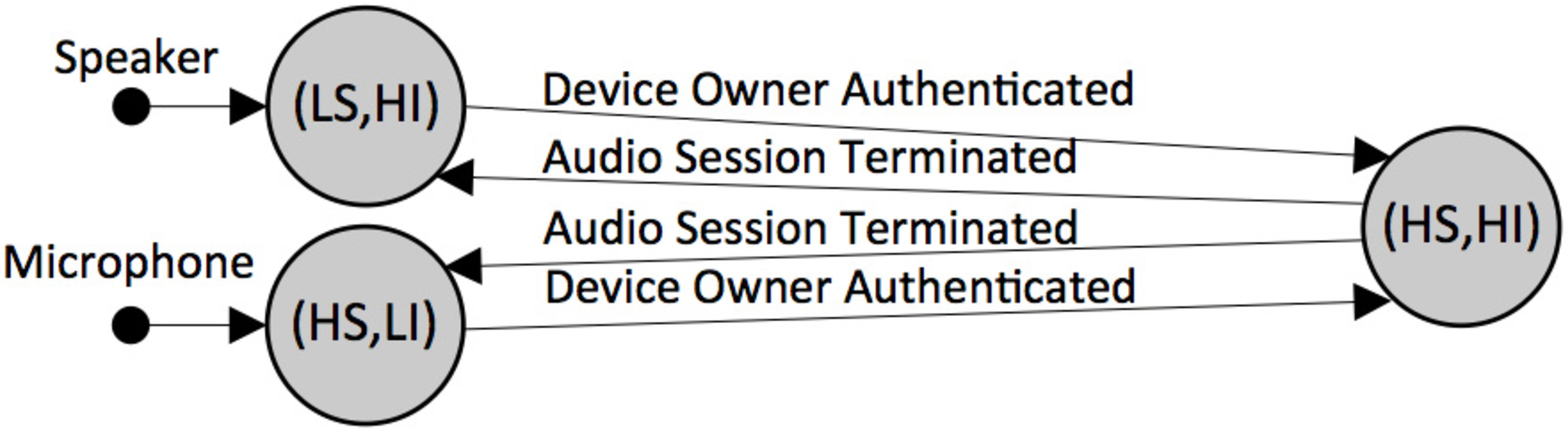}
\caption{Security Levels for External Parties.}
\label{fig:statemac}
\end{figure}

By default, if an internal party is using the speaker, \system identifies the security levels of an external party listening to the speaker as low-secrecy and high-integrity ($L3$ = \texttt{(LS,HI)}). This configuration prevents a system app or service from leaking security-sensitive information to an external entity different from the device owner, and at the same time prevents market apps from producing audio that would mislead the device owner or affect an external, in-proximity device.
On the other hand, if the internal party is using the microphone, \system identifies the security levels of an external party sending input to the microphone as high-secrecy and low-integrity ($L4$ = \texttt{(HS,LI)}). This configuration prevents a market app from eavesdropping the external party, and at the same time prevents any system app and service from receiving voice commands from an external party different from the device owner.
In both cases, the security levels of both external parties  are elevated to high-secrecy and high-integrity ($L3 = L4$ = \texttt{(HS,HI)}) upon authentication of the owner of the device, as shown in Figure~\ref{fig:statemac}. User authentication is an orthogonal problem to out research objective. We assume there exists a authentication mechanism. In the evaluation (Section \ref{evaluation}), we use screen lock passcodes to authenticate device owners. Exploring more suitable authentication mechanisms is future work.

\subsection{Prevention of Unsafe Information Flows} \label{mls}

\begin{figure}[h]
\centering
\includegraphics[width=80mm]{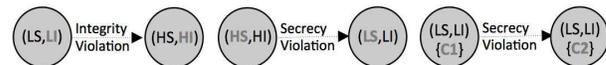}
\caption{Unsafe Audio Flows.}
\label{fig:unsafe}
\end{figure}

\system considers an audio channel's information flow ({\em audio flow}) {\em unsafe} if produced by a low-integrity party and directed to a high-integrity party, such as a flow from a market app to a system service. Similarly, \system considers an audio flow {\em unsafe} if produced by a high-secrecy party and directed to a low-secrecy party, such as a flow from the device owner to a market app. Finally, audio flows between low-secrecy low-integrity parties (apps) are also considered unsafe by \system, which separate apps by assigning them to different categories. Unsafe audio flows are shown in Figure~\ref{fig:unsafe}, where \textbf{C1} and \textbf{C2} are categories assigned to apps.

\vspace{-0.08in}

\vspace{0.08in}

In \system, attacks based on audio channels are prevented by identifying and blocking unsafe information flows by enforcing MLS access control policies, such as Biba~\cite{Biba} and Bell-LaPadula (BLP)~\cite{blp}. Consequently, \system prevents any audio produced by a low-integrity party from flowing to a high-integrity party, and any audio produced by a high-secrecy party from flowing to a low-secrecy party. Additionally, to avoid unsafe flows among apps, \system prevents the creation of audio channels between low-secrecy low-integrity processes (apps) by leveraging the use of categories. 

Extending the SELinux reference monitor is challenging because \system has to ensure {\em complete mediation} of all security-sensitive operations~\cite{monitor}.  For \system, we must identify the right locations to place its {\em Audio Hooks} to mediate every access to the microphone and speaker by any process. There are two possible alternatives: inside the Android framework/middleware or inside the Linux kernel. To achieve complete mediation of accesses to the microphone and speaker, which are low-level system resources, kernel mediation would seem to be most appropriate. Unfortunately, hooks inside the kernel do not have visibility into the actual processes that are requesting access to these resources. This is due to the fact that accesses to system resources are always performed by system services ({\em Media Server}) on behalf of the some requesting process, the one running an app or another system service. For the microphone and speaker, the {\em Media Server} provides complete mediation, as shown in Figure~\ref{fig:arch}, because it is the only system service allowed to access the microphone and speaker device files, due to specific MAC rules specified by SE Android \cite{seandroid}. Therefore, we extend the SELinux reference monitor by placing hooks in the {\em Media Server} in the Android framework/middleware.  

\subsection{Resolution of Unsafe Audio Flows} \label{resolver}

Blocking every audio flow from a high-secrecy party to a low-secrecy party would prevent system apps and services from performing some expected operations, such as producing a ring tone on an incoming call. Similarly, blocking every audio flow from a low-integrity party to a high-integrity party would prevent market apps from performing some expected operations, such as producing sound when a message is received. To preserve these functional requirements, \system uses {\em resolvers} to enable privileged processes to resolve some information flow errors.

\begin{figure}[h]
\centering
\includegraphics[width=70mm]{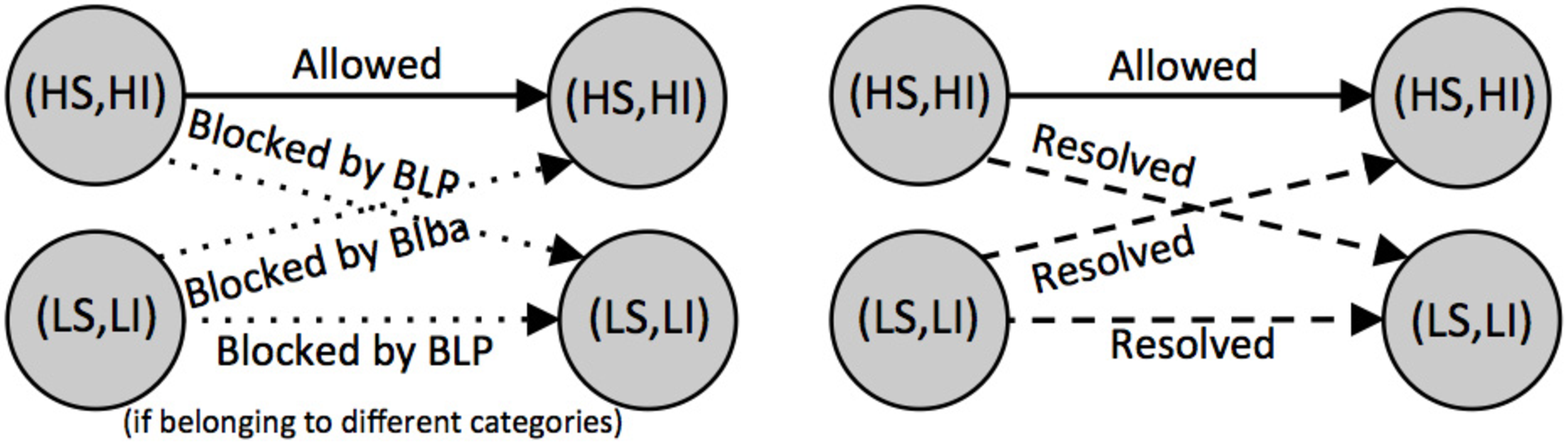}
\caption{Audio Flows Allowed and Resolved by \system}
\label{fig:flows}
\end{figure}

Figure~\ref{fig:flows} shows the information flows in need of resolution. To implement information flow error resolution, \system uses a callback mechanism, as in the {\em Android Security Modules} (ASM) \cite{Heuser} framework, to notify the system app or service at risk about the information flow error.  The system app or service is provided with the identity of one or more {\em resolvers}, known methods for preventing information flow errors.  We examine different types of resolvers in Section~\ref{evaluation}, but one example would be to only play approved audio files (e.g., ring tones) that do not contain secrets or malice.  Unlike ASM, only system apps and services may receive callbacks and only those currently using audio channels are notified.

\subsection{Resolution of Flows to External Parties} \label{trusted}

Resolvers can automatically resolve information flow errors to/from speakers and microphones.  However, some information flow errors may only be resolved by the external party, i.e., the device owner.  For example, if the device owner uses a market app that uses the microphone (and has sufficient Android permissions), it is not possible for a system app or service to ensure that the device owner is safe from eavesdropping.  

To enable resolution, \system creates {\em trusted paths} \cite{orangebook} between the device owner and \system, by implementing mechanism similar to those supported by {\em User-Driven Access Control} (UDAC) ~\cite{Roesner}.  In UDAC, each app may choose a trusted module to run when user interaction is required.  The trusted module can also convey the results of the user interaction to other trusted components, such as \system.  Unfortunately, to use UDAC, each app must leverage trusted modules when user interaction is needed, but such modules are not currently deployed and app developers may fail to use UDAC even when such modules become available. 
Instead, in the construction of \system, we modify the {\em Media Server} to recognize when there remains an unsafe information flow to external parties (after applying resolvers), so it can apply  a trusted module to notify and gain approval from users.  Therefore, \system does not depend on untrusted app developers. The user approval cannot change the access control policy in a discretionary manner, it only changes the labeling of the external party for the communication channel. An event cache is used to automatically resolve identical information flows happening in a short time interval, which reduces the resolution overhead and avoids bothering the user. 

\begin{figure}[h]
\centering
\includegraphics[width=70mm]{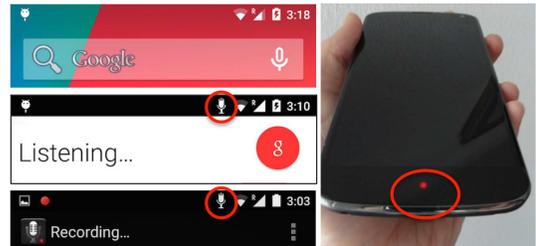}
\caption{Microphone Icon and Notification Light.}
\label{fig:notifications}
\end{figure}

\system provides two mechanisms for the creation of trust-ed paths. First, whenever a low-integrity and low-secrecy party (i.e., a market app) asks for access to the microphone, \system requires approval by the device owner. The communication is not authorized until the user allows recording through the trusted path. \system provides the user with the information about the parties that would communicate through the creation of the audio channel if the access request is granted. 
Second, \system notifies the user as long as the microphone is in use. \system uses a microphone icon on the status bar and a notification light on the front side of the device to notify the user that the microphone is being accessed, as shown in Figure~\ref{fig:notifications}. 
The notification light replaces the microphone icon as the notification mechanism as soon as the screen goes off. 

A recent user study found that visual notification catches the user attention for between 64-77\% of the cases~\cite{Bianchi}. 

\vspace{-0.05in}

\section{Implementation of {\secit \large \system}}
In this section, we provide details about the implementation of the \system Framework based on a vanilla Android OS (version 5.0.1\_r1), obtained from the official \emph{Android Open Source Project} (AOSP) repository \cite{aosp}. The \system prototype has been tested on a Nexus 5 smartphone. The \system source code and the code for the apps implementing attacks (described in this paper) will be made available on \texttt{https://github.com/gxp18/AuDroid}.
The current \system prototype is implemented in less than 520 LOC in C++, less than 130 LOC in C, and about 120 LOC in Java. Consequently, the impact in terms of the customization needed to integrate \system in a vanilla Android OS distribution is negligible. We have written a simple patch that automatically integrates the additional modules into the vanilla distribution.

\begin{figure}[h]
\centering
\includegraphics[width=70mm]{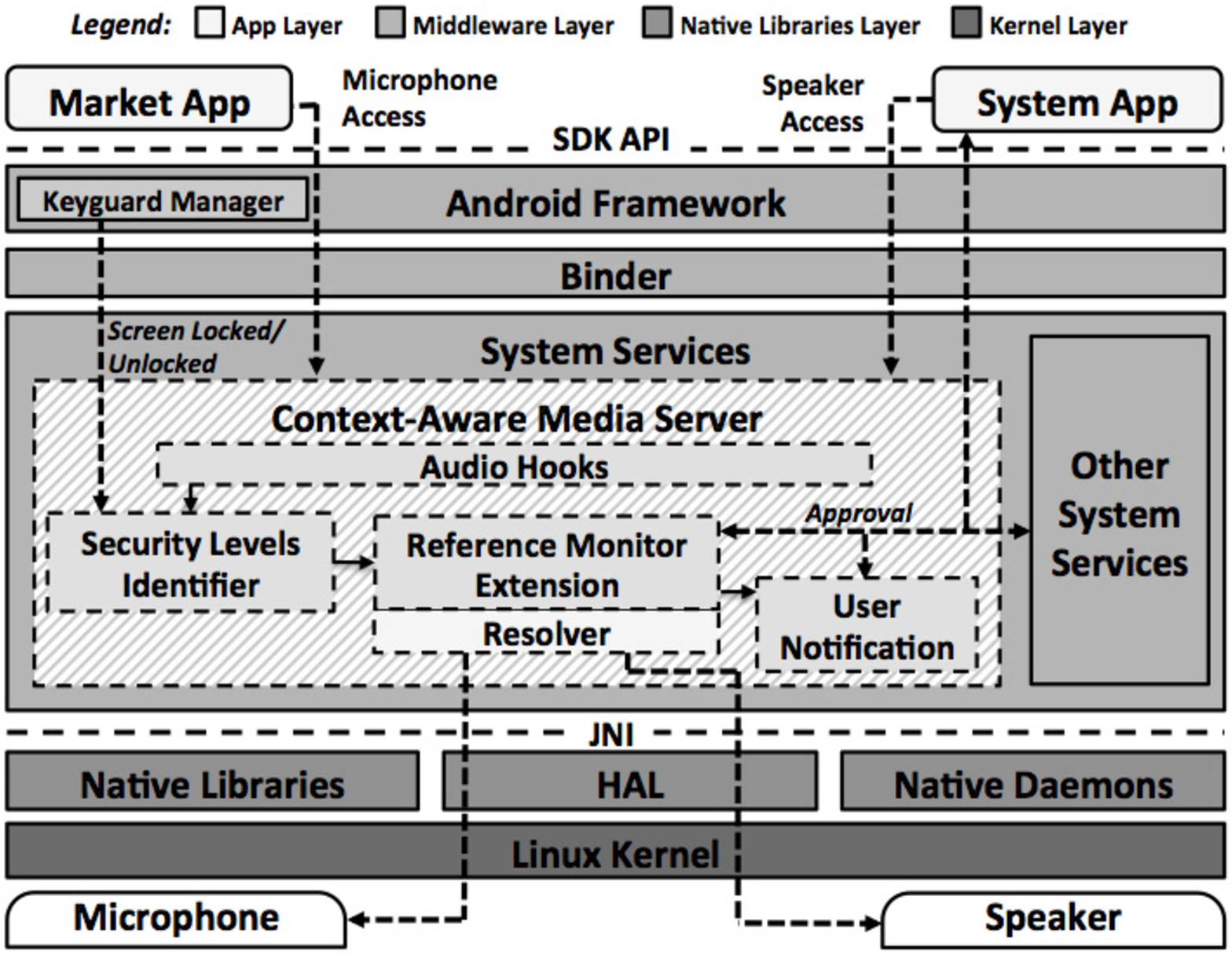}
\caption{\system Framework Architecture.}
\label{fig:implementation}
\end{figure}

An overview of the \system Framework architecture is depicted in Figure \ref{fig:implementation}. The difference between \system and the original Android Framework architecture (Figure~\ref{fig:arch}) is at the System Services layer. In \system, the \emph{Media Server} is made context-aware by integrating it with the following new additional modules: \emph{Audio Hooks}, \emph{Security Level Identifier}, \emph{Reference Monitor Extension}, and \emph{User Notification} modules. Additional modules are shown as dashed, light-grey boxes in Figure \ref{fig:implementation}.

\subsection{Audio Hooks} \label{hooks}

\system makes use of four \emph{Audio Hooks} to mediate access to the microphone and speaker. It uses two hooks for each system resource in order to capture the resource acquisition and release. \emph{Audio Hooks} are placed inside the \emph{Audio System} module, part of the \emph{Media Server}, and located at the following four mediation points: \texttt{AudioSystem::startInput()}, \texttt{AudioSystem::stopInput()}, \texttt{AudioSystem::startOutput()} and \texttt{AudioSystem::stopOutput()}. These are the only entry and exit points for obtaining the microphone and speaker for any apps or other system services, as validated through dynamic analysis of the Android framework source code and SDK API. Each hook retrieves the PID of the calling process which is passed, together with the mediation point, as parameter in a call to the \emph{Security Levels Identifier} module. A code snippet of an {\em Audio Hook} is reported below:

\lstset{
numbers=left, 
numberstyle=\tiny, 
numbersep=3pt,  
  basicstyle=\small\normalfont\sffamily,    
  stepnumber=1,                           
  numbersep=3pt,                         
  tabsize=1,                              
  extendedchars=true,                     %
  breaklines=true,                        
  captionpos=t,                           
  mathescape=true,
  stringstyle=\color{white}\ttfamily, 
  showspaces=false,           
  showtabs=false,             
  xleftmargin=4pt,
  framexleftmargin=3pt,
  framexrightmargin=0pt,
  framexbottommargin=0pt,
  framextopmargin=0pt,
  showstringspaces=false      %
}

 \begin{lstlisting} [basicstyle=\scriptsize \ttfamily ]
pid_t pid = IPCThreadState::self()->getCallingPid();
int sec_l = SecurityLevelIdentifier::getLevels(pid); 
PolicyEnforcer::enforce(sec_l, mediationPoint);
 
\end{lstlisting}

Since SELinux restricts access to the microphone or speak-er only to the {\em Media Server} and the {\em Audio Hooks}, complete mediation by the {\em Media Server} would be sufficient to prevent circumvention. We used dynamic analysis, of the Android Framework source code and SDK API, to validate complete mediation, checking that every access to the microphone or speaker was accompanied by an invocation of an appropriate {\em Audio Hook}.  Retaining such logging could be used to detect errors, if any exist.

\subsection{Security Levels Identifier}

The \emph{Security Levels Identifier} is implemented in C++ and uses the PID of the calling process to categorize processes accessing the microphone or speaker, as described in Section \ref{levels}. The \emph{Security Levels Identifier} categorizes processes related to market apps as low-secrecy and low-integrity, and processes related to system apps or services as high-secrecy and high-integrity, in accordance with our security model. Furthermore, the \emph{Security Levels Identifier} uses information provided by the {\em Keyguard Manager}, a module inside the Android framework, to determine if the device screen is locked or unlocked as optional mechanism to authenticate the device owner, as described in Section \ref{levels}.
\vspace{-0.08in}

\subsection{Reference Monitor Extension}

The \emph{Reference Monitor Extension} is a software module, implemented in C++, in charge of enforcing the access control policy in response to calls coming from the \emph{Audio Hooks}. The \emph{Reference Monitor Extension} is context-aware, in that, it uses the security levels identified by the {\em Security Levels Identifier} while authorizing access to the audio channels produce at runtime.
Additionally, the {\em Reference Monitor Extension} can leverage resolvers whenever an information flow violation is identified, as described in Section \ref{resolver}. Whenever a resolver is involved, the {\em Reference Monitor Extension} uses the callback mechanism to obtain approval from the system app or service at risk.
Finally, the {\em Reference Monitor Extension} builds trusted paths with the device owner to obtain approval for the use of the microphone by low-secrecy low-integrity processes. Upon approval, the access to the microphone is made visible to the user through the notification mechanism, as described in Section \ref{trusted}.
\vspace{-0.08in}

\subsection{User Notification} 

In \system, the device owner is notified about accesses to the microphone, by running processes, via two notification mechanisms. The first one is the Microphone Icon that appears on the status bar anytime a process accesses the microphone after being authorized by the {\em Media Server}. An icon depicting a microphone is shown on the device system status bar for the entire audio session, as depicted in Figure \ref{fig:notifications}. By tapping on the microphone icon the device owner obtains further information about the app currently accessing the microphone. The second notification consists of a blinking notification light, on the front side of the mobile device, activated whenever the screen goes off if a process is accessing the microphone after being authorized by the {\em Media Server}. These two notification mechanisms increase the device owner awareness anytime the microphone is accessed. We rely on the user common sense in avoiding speaking about personal matters or sensitive information whenever access to the microphone is signaled.

\vspace{-0.05in}

\section{Evaluation} \label{evaluation}
In this section, we evaluate the effectiveness of \system in preventing the six types of attack scenarios for audio channels discussed in Section \ref{attacks}, examine whether 17 widely-used apps and services can still be run effectively under \system, and measure the performance impact of \system.  
\vspace{-0.08in}

\subsection{Attack Prevention}

Table~\ref{table:attack} compares the effectiveness of \system in preventing the six types of attack scenarios on audio channels outlined in Section~\ref{attacks} to Android, other related work described in Section~\ref{related}, and Simple Isolation\footnote{No simultaneous access to the microphone and speaker by two different processes.}.  \system is capable of preventing all six types of attack scenarios.  Other defenses prevent no more than two types of these attack scenarios because they lack awareness of the impact of external parties.
Further, other defenses are of limited applicability or may often cause false positives, as described in the next section.

\begin{table}[t]
\centering
\caption{Attack Prevention Analysis} 

\tiny

\label{table:attack}
\tabcolsep=0.07cm
\vspace*{+2pt}

\begin{tabular}{l|c|c|c|c|c|c|} 
\cline{2-7} 
\multicolumn{1}{c|}{\multirow{9}{*}{\begin{tabular}[c]{@{}c@{}}\textit{Legend:} \\\\ \cmark Attack Prevented \\ \xmark Attack Succeeded \\  $\triangle$ Attack might \\ be Prevented \\  \end{tabular}}}& 

\multicolumn{1}{c|}{\multirow{9}{*}{ \begin{turn}{90} \hspace{-2mm} \begin{tabular}{@{}c@{}}\emph{Touchless Control} \\ Type 1\\Scenario 1\end{tabular} \end{turn} }} & 
\multicolumn{1}{c|}{\multirow{9}{*}{\begin{turn}{90}  \begin{tabular}{@{}c@{}}\hspace{-3mm}\emph{Keylogger} \\ Type 1\\Scenario 2 \end{tabular}\end{turn} }}& 
\multicolumn{1}{c|}{\multirow{9}{*}{\begin{turn}{90} \hspace{-2mm} \begin{tabular}{@{}c@{}}\emph{Device Control} \\ Type 2 \\ Scenario 3 \end{tabular} \end{turn} }} &
\multicolumn{1}{c|}{\multirow{9}{*}{\begin{turn}{90} \hspace{-1mm}\begin{tabular}{@{}c@{}}\emph{Speak Out} \\ Type 2 \\Scenario 4 \end{tabular} \end{turn} }}& 
\multicolumn{1}{c|}{\multirow{9}{*}{\begin{turn}{90}\hspace{-2mm} \begin{tabular}{@{}c@{}}\emph{Voice Commands} \\ Type 3 \\Scenario 5 \end{tabular} \end{turn}}}& 
\multicolumn{1}{c|}{\multirow{9}{*}{\begin{turn}{90}\hspace{-2mm} \begin{tabular}{@{}c@{}}\emph{Stealthy Recording  } \\ Type 3 \\Scenario 6 \end{tabular} \end{turn}}}\\  

&& & & & & \\  
&& & & & & \\  
&& & & & & \\  
&& & & & & \\ 
&& & & & & \\  

&& & & & & \\ 
&& & & & & \\ 
&& & & & & \\ \hline  

\rowcolor{Gray}\multicolumn{1}{|c|}{ \texttt{Base Android}}  &  \xmark & \xmark  & \xmark &\xmark&\xmark&\xmark \\ \hline   \hline
\rowcolor{Gray1}\multicolumn{1}{|c|}{\texttt{Simple Isolation}}&  \cmark & \cmark  & \xmark &\xmark & \xmark & \xmark \\ \hline

\rowcolor{Gray2}\multicolumn{1}{|c|}{\texttt{\system}}& \cmark & \cmark & \cmark  & \cmark & \cmark & \cmark \\ \hline  \hline
\multicolumn{1}{|c|}{\texttt{\cellcolor[gray]{0.8}\begin{tabular}{@{}c@{}}Google Voice Search bug fix \end{tabular} }}& \cmark& \xmark  & \xmark &\xmark&\xmark&\xmark\\ \hline
\multicolumn{1}{|c|}{\texttt{\cellcolor[gray]{0.8}\begin{tabular}{@{}c@{}}Control Access to Speaker\end{tabular}} }& \cmark & \xmark & \xmark &\xmark & \xmark & \xmark  \\ \hline
\multicolumn{1}{|c|}{\texttt{\cellcolor[gray]{0.8}\begin{tabular}{@{}c@{}}System Services Permission\end{tabular}} } & $\triangle$  & $\triangle$  & \xmark &\xmark&\xmark&\xmark \\ \hline
\multicolumn{1}{|c|}{\cellcolor[gray]{0.8}\texttt{\begin{tabular}{@{}c@{}}Voiceprint Recognition\end{tabular}}}  & \cmark & \xmark & \xmark & \xmark & \cmark & \xmark \\ \hline 
\end{tabular}
\vspace*{-\baselineskip}
\end{table}

\textbf{Touchless Control} (Table \ref{table:attack} column 1) refers to the attack where a malicious app makes use of an audio channel of Type 1 to exploit a system service receiving voice commands to perform security-sensitive operations \cite{Jang,Diao}. This type of attack is prevented by \texttt{Simple Isolation}, because two processes cannot access - at the same time - the microphone and speaker. Similarly, \system detects an unsafe flow from a low-integrity process (malicious app) to a high-integrity process (system service). Rows 4-7 show that other defense mechanisms can prevent this attack. In particular, the $\triangle$ symbol (in row 6) highlights that, a permission mechanism used to authorize the use of System Services could prevent the attack, if the user does not grant the permissions to use the {\em Media Server} for a malicious app.

\textbf{Keylogger} (Table \ref{table:attack} column 2) refers to an attack where a malicious app uses an audio channel of Type 1 to eavesdrop the password typed by the device owner and spoken out by the {\em TalkBack} accessibility service \cite{Jang}. \texttt{Simple Isolation} prevents the attack because two processes cannot access the microphone an speaker at the same time. \system prevents the attack because a unsafe flow from a high-secrecy process (accessibility service) to a low-secrecy process (malicious app) is detected. Furthermore, the $\triangle$ symbol (in row 6) highlights the fact that a permission mechanism, used to authorize the use of System Services, could prevent the attack if the user does not grant the malicious app permission to use the {\em Media Server}. 

\textbf{Device Control} (Table \ref{table:attack} column 3) is an attack performed by using a malicious app running on a device as a source of malicious voice commands (such as those reported in Table \ref{table:commands}) to attack another nearby device. The attack is performed by using an audio channel of Type 2. \texttt{Simple Isolation} does not prevent the attack because it does not consider audio channels involving external parties. On the other hand, \system detects the unsafe flow from a low-integrity process to a high-integrity external party. 

\textbf{Speak Out} (Table \ref{table:attack} column 4) refers to a malicious app eavesdropping voice and sound through the device's microphone to collect security-sensitive information, such as private conversations, successively leaked to an adversary through the device's speaker as soon as the device owner, victim of the attack, is away from the device. The adversary makes use of an audio channel of Type 2 to bypass any lock screen protection mechanism. \texttt{Simple Isolation} does not prevent the attack because it does not consider audio channels involving external parties. On the other hand, \system detects the unsafe flow from a low-integrity process to a high-integrity external party. Furthermore, \system prevents a malicious app from eavesdropping on the device owner because the user approval is required before a low-secrecy process can access the microphone.

\textbf{Voice Commands} (Table \ref{table:attack} column 5) is an attack performed by an adversary directly interacting with the target device via malicious voice commands. In this attack, the adversary uses an audio channel of Type 3. \texttt{Simple Isolation} does not prevent the attack because it does not consider audio channels involving external parties. On the other hand, \system prevents a malicious user, different from the device owner, from delivering voice commands to a system service, by identifying the external party (user) as low-integrity and the system service as high-integrity. As shown in row 7, \texttt{Voiceprint Recognition} can prevent the attack, unless the adversary replays recorded device owner voice commands.

\textbf{Stealthy Recording} (Table \ref{table:attack} column 6) refers to an attack where a malicious app uses an audio channel of Type 3 to stealthily record audio through the device's microphone in order to eavesdrop the device owner and the surrounding environment. \system prevents a malicious app from eavesdropping the device owner voice because the user approval is required before a low-secrecy process can access the device's microphone.

\vspace{-0.08in}

\subsection{System Functionality}

\begin{table*}

\centering
\caption{System Functionality Analysis}

\tiny
\label{table:functionality}
\tabcolsep=0.08cm
\vspace*{+2pt}

\begin{tabular} {l|c|c|c|c|c|c|c|c|c|c|c|c|c|c|c|c|c|}  
\cline{2-18}
\multicolumn{1}{c|}{\multirow{6}{*}{\begin{tabular}[c]{@{}c@{}}  \textit{Legend:} \\  \\\cmark App Runs \\ \texttt{IV} Integrity Violation \\ \texttt{SV} Secrecy Violation \\ \texttt{SIV} Secrecy and Integrity \\ Violation \end{tabular}}}
&  \multicolumn{7}{c|}{\cellcolor[gray]{0.9}System Apps}& \multicolumn{10}{c|}{  \cellcolor[gray]{0.6} Market Apps }\\ 
\cline{2-18} 
& \multirow{5}{*}{ \begin{turn}{90}Voice Dialer \hspace{2.5mm} \end{turn} } 
& \multirow{5}{*}{ \begin{turn}{90}Music \hspace{9mm} \end{turn} }
& \multirow{5}{*}{ \begin{turn}{90}Voice Search \hspace{2mm} \end{turn} }
& \multirow{5}{*}{ \begin{turn}{90}Phone \hspace{9mm} \end{turn} }
& \multirow{5}{*}{ \begin{turn}{90}Hangouts \hspace{5.5mm} \end{turn} }
& \multirow{5}{*}{ \begin{turn}{90}Browser \hspace{7mm} \end{turn}  } 
& \multirow{5}{*}{ \begin{turn}{90}Maps \hspace{9.5mm} \end{turn}  }
& \multirow{5}{*}{ \begin{turn}{90}Pandora \hspace{6.5mm} \end{turn}  }
& \multirow{5}{*}{ \begin{turn}{90}Spotify \hspace{7.5mm} \end{turn} }
& \multirow{5}{*}{ \begin{turn}{90}Viber \hspace{9.5mm} \end{turn}  }
& \multirow{5}{*}{ \begin{turn}{90}WhatsApp \hspace{4mm} \end{turn}  } 
& \multirow{5}{*}{ \begin{turn}{90}Snapchat \hspace{5.5mm} \end{turn}  }
& \multirow{5}{*}{ \begin{turn}{90}Facebook \hspace{5.5mm} \end{turn} }
& \multirow{5}{*}{ \begin{turn}{90}Skype \hspace{9mm} \end{turn}  } 
& \multirow{5}{*}{ \begin{turn}{90}Voice Memos \hspace{1.5mm} \end{turn}  }
& \multirow{5}{*}{ \begin{turn}{90}Voice Recorder \hspace{-0.5mm} \end{turn}  }
& \multirow{5}{*}{ \begin{turn}{90}Call Recorder \hspace{1mm} \end{turn}  } \\  
& & & & & & & & & & & & & & & & & \\  
& & & & & & & & & & & & & & & & & \\  
& & & & & & & & & & & & & & & & & \\   
& & & & & & & & & & & & & & & & & \\  
& & & & & & & & & & & & & & & & & \\   
& & & & & & & & & & & & & & & & & \\   
& & & & & & & & & & & & & & & & & \\   \hline  

\multicolumn{1}{|l|}{\cellcolor[gray]{0.8}\texttt{Simple Isolation}} & \cmark & \cmark  & \cmark & \cmark & \cmark & \cmark & \cmark & \cmark & \cmark & \cmark & \cmark & \cmark  & \cmark & \cmark & \cmark & \cmark & \cmark \\ \hline  

\multicolumn{1}{|l|}{\cellcolor[gray]{0.8}\system \texttt{MLS}}  & \cmark & \cmark  & \cmark & \cellcolor[gray]{0.6}\texttt{SV} & \cellcolor[gray]{0.6}\texttt{SV} & \cmark & \cmark & \cellcolor[gray]{0.8}\texttt{IV} & \cellcolor[gray]{0.8}\texttt{IV} & \cellcolor[gray]{0.4}\texttt{SIV}  &\cellcolor[gray]{0.4}\texttt{SIV} & \cellcolor[gray]{0.4}\texttt{SIV}  &\cellcolor[gray]{0.4}\texttt{SIV} &\cellcolor[gray]{0.4}\texttt{SIV} &\cellcolor[gray]{0.4}\texttt{SIV} & \cellcolor[gray]{0.4}\texttt{SIV} & \cellcolor[gray]{0.4}\texttt{SIV}  \\ \hline  

 \multicolumn{1}{|l|}{\cellcolor[gray]{0.8}\system \texttt{User Approval}}& \cmark  & \cmark  &  \cmark  & \cellcolor[gray]{0.6}\texttt{SV} & \cellcolor[gray]{0.6}\texttt{SV} & \cmark  & \cmark  &  \cellcolor[gray]{0.8}\texttt{IV} &  \cellcolor[gray]{0.8}\texttt{IV} &  \cellcolor[gray]{0.8}\texttt{IV}  &   \cellcolor[gray]{0.8}\texttt{IV} &  \cellcolor[gray]{0.8}\texttt{IV} &  \cellcolor[gray]{0.8} \texttt{IV} &  \cellcolor[gray]{0.8}\texttt{IV} &  \cellcolor[gray]{0.8} \texttt{IV} & \cellcolor[gray]{0.8} \texttt{IV} &  \cellcolor[gray]{0.8}\texttt{IV} \\ \hline 

\multicolumn{1}{|l|}{\cellcolor[gray]{0.8}\system \texttt{Resolver 1}} &\cmark  & \cmark  &  \cmark  & \cmark & \cmark & \cmark & \cmark & \cellcolor[gray]{0.8}\texttt{IV} & \cellcolor[gray]{0.8}\texttt{IV} & \cellcolor[gray]{0.4}\texttt{SIV}  &\cellcolor[gray]{0.4}\texttt{SIV} & \cellcolor[gray]{0.4}\texttt{SIV}  &\cellcolor[gray]{0.4}\texttt{SIV} &\cellcolor[gray]{0.4}\texttt{SIV} &\cellcolor[gray]{0.4}\texttt{SIV} & \cellcolor[gray]{0.4}\texttt{SIV} & \cellcolor[gray]{0.4}\texttt{SIV}\\ \hline 

\multicolumn{1}{|l|}{\cellcolor[gray]{0.8}\system \texttt{Resolver 2}} &\cmark  & \cmark  &  \cmark  & \cellcolor[gray]{0.6}\texttt{SV} & \cellcolor[gray]{0.6}\texttt{SV} & \cmark & \cmark & \cmark & \cmark & \cellcolor[gray]{0.6}\texttt{SV} & \cellcolor[gray]{0.6}\texttt{SV}  & \cellcolor[gray]{0.6}\texttt{SV} & \cellcolor[gray]{0.6}\texttt{SV} & \cellcolor[gray]{0.6}\texttt{SV} & \cellcolor[gray]{0.6}\texttt{SV} & \cellcolor[gray]{0.6}\texttt{SV} & \cellcolor[gray]{0.6}\texttt{SV}\\ \hline

\multicolumn{1}{|l|}{\cellcolor[gray]{0.8}\system} & \cmark & \cmark  & \cmark & \cmark & \cmark & \cmark & \cmark & \cmark & \cmark & \cmark & \cmark & \cmark  & \cmark & \cmark & \cmark & \cmark & \cmark \\ \hline  

\hline

\multicolumn{1}{|l|}{  \cellcolor[gray]{0.6}Requested User Approval}& \cellcolor[gray]{0.8}no & \cellcolor[gray]{0.8}no & \cellcolor[gray]{0.8}no  & \cellcolor[gray]{0.8}no & \cellcolor[gray]{0.8}no & \cellcolor[gray]{0.8}no & \cellcolor[gray]{0.8}no & \cellcolor[gray]{0.8}no & \cellcolor[gray]{0.8}no & yes  & yes  & yes &  yes & yes &  yes & yes & yes \\ \hline 

\multicolumn{1}{|l|}{  \cellcolor[gray]{0.6}User Notified}& yes & \cellcolor[gray]{0.8}no & yes  & yes & yes & yes & yes & \cellcolor[gray]{0.8}no & \cellcolor[gray]{0.8}no & yes  & yes  & yes &  yes & yes &  yes & yes & yes \\ \hline

\multicolumn{1}{|l|}{  \cellcolor[gray]{0.6}App uses Microphone}& yes & \cellcolor[gray]{0.8}no & yes  & yes & yes & yes & yes & \cellcolor[gray]{0.8}no & \cellcolor[gray]{0.8}no & yes  & yes  & yes &  yes & yes &  yes & yes & yes \\ \hline
 
\multicolumn{1}{|l|}{  \cellcolor[gray]{0.6}App uses Speaker}& yes & yes & yes  & yes & yes & yes & yes & yes & yes & yes  & yes  & yes &  yes & yes &  yes & yes & yes \\ \hline

\end{tabular}
\quad
\begin{tabular}{l}
\tiny 1  Snapchat (Take Video)\\
\tiny 2  Facebook (Send Voice Message)\\
\tiny 3  Whatsapp (Send Voice Message)\\
\tiny 4  Voice Recorder (Send Voice Message)\\
\tiny 5  Viber (Send Voice Message) \\
\tiny 6  Google Voice Search (Voice Search)\\
\tiny 7  Browser (Watch Video)\\
\tiny 8  Skype (Video Call)\\
\tiny 9  Call Recorder (Record Phone Call)\\
\tiny 10 Pandora (Listen to a Song)\\
\tiny 11 Spotify (Listen to a Song)\\
\end{tabular}

\vspace*{-\baselineskip}
\end{table*}

We next evaluate the impact of \system on the ability of apps and services to operate normally. 
The results of our analysis are reported in Table \ref{table:functionality}.  We evaluate \system for 10 market apps and 7 system apps distributed with the Android OS from Google. We select market apps that are among the most popular apps available on Google Play.  We also choose market and system apps that use either the speaker or microphone or both, as indicated by the last four rows in Table~\ref{table:functionality}.

Row 1 in Table \ref{table:functionality} shows that, by enforcing {\em Simple Isolation}, all the system and market apps would work fine although interaction among apps is not allowed. For example, the user cannot use the Voice Recorder app to tape the music produced by the Music app. Therefore, although the system functionality is preserved, there is an indirect impact on how apps can interact. 

We then analyze the impact of using \system when MLS is applied to enforce Biba and Bell-LaPadula. From row 2 in Table \ref{table:functionality}, we observe that two security violations are reported for the Phone and Hangouts system apps, which are due to the fact that these apps produce a sound on incoming calls or message receptions, even when the external party is identified as low-secrecy (i.e. device screen locked). This is seen by \system as a flow from a high-secrecy party (system app) to a low-secrecy party (user different from the device owner). Furthermore, from row 2, we observe two integrity violations in correspondence of Pandora and Spotify. This is due to the fact that these apps access the speaker to produce music and sounds, therefore \system sees a flow from a low-integrity party to a external high-integrity party. Finally, row 2 reports secrecy and integrity violations for the remaining market apps from Viber to Call Recorder. The integrity violations are due to the same reason explained for Pandora and Spotify, whereas the secrecy violations are due to the fact that \system sees a flow from an external high-secrecy party (i.e. device owner) to a low-secrecy party (market app). 

Row 3 shows how \system resolves the secrecy violations relative to the market apps by using the user approval mechanism. In particular, whenever a market app uses the microphone, the device owner is notified and can approve or deny the access.

Row 4 shows the effect of using a resolver (\texttt{Resolver 1}) that allows system apps, like Phone and Hangouts, to play approved ring tones and notification sounds even when the external party is identified as low-secrecy. 

Row 5 shows the effect of using a second resolver (\texttt{Resolver 2}) to allow market apps, like Pandora and Viber, to play approved audio files that do not contain malice, such as ring tones, notification sounds and sound tracks, even when the external party is identified as high-integrity.

By combining the user approval and the resolvers, \system runs all the tested apps correctly without impact on the system functionality, as show in row 6.

\subsection{Performance Overhead}

Existing benchmarks for Android, such as AM-Bench \cite{am-bench} and Android Workload Suite \cite{aws}, do not target apps making extensive use of the microphone/speaker and are not publicly available. Instead, we measured the performance overhead  introduced by \system through the following 3 experiments on a Nexus 5 running Android aosp-5.0.1\_r1. 

We first measured the overhead introduced by \system while handling each access request for the microphone and speaker, by measuring the time interval from the time the request is made by the process running the app to the time the access is granted/denied by the \emph{Media Server}. We found that, on average over 10,000 requests, the original Android system required 20.35$\pm$1.90$\mu$s to handle an access request for the speaker and 25.36$\pm$2.01$\mu$s for the microphone. When \system is activated to mediate accesses to the microphone and speaker, each access request for the speaker is handled in 24.47$\pm$1.86$\mu$s and 30.11$\pm$1.99$\mu$s for the microphone. The main reason for the overhead introduced by \system was the time necessary to recreate the context (i.e., PIDs of involved processes) used to make the access control decision. Note that enabling the \system notification mechanism (icon microphone and notification light) increased the access time for the microphone to 38.43$\pm$2.11$\mu$s.

Second, we measured the overhead of running a sequence of 11 well-known apps (listed on the left side of Table \ref{table:functionality}) that create audio sessions, closing each app after running a 30s session before opening the subsequent one. We found that both, the original Android system and \system, took 591$\pm$21.93s to complete on average.  Unfortunately, such a measurement required a human in the loop, but the audio session length reduced the variability caused by human actions.  The main overhead for \system incurred on opening and closing speaker and microphone connections, which are infrequent and low overhead operations.

Third, we used the Voice Recorder app to examine the overhead if we tried to create audio sessions as quickly as humanly possible on the device.  To do this, the user continuously tapped on the microphone button, as quickly as possible, to keep recording new voice messages. The experiment was repeated 5 times over a duration of approximately one minute. We registered a maximum of 21 access requests/minute for the microphone and a maximum of 53 access requests/minute for the speaker. We found that, with the highest possible number of audio sessions in a minute, both, the original Android and \system completed the task in approximately 59.54s, indicating overhead not detectable.

\vspace{-0.05in}

\section{Related Work} \label{related}
Diao \emph{et al.}~\cite{Diao} discuss how a missing check in the  \emph{Google Voice Search} (GVS) app allows any app to send an intent and activate GVS app in \emph{Voice Dialer Mode}, a mode that unlocks the screen and allows dialing phone numbers through voice commands. They propose enabling the GVS app to suspend any other process trying to access the speaker while it is actively listening for voice commands, ensuring that GVS app receives voice commands only from the device owner and not from audio played by third processes. The two defense mechanisms discussed above are \emph{ad hoc} solutions for the GVS app, in that, they do not prevent attacks against other apps or system services. They also suggest that voiceprint recognition techniques \cite{Beigi} could be adopted to verify the identity of the user speaking to the device, by using biometrics voice characteristics. 
However, an adversary can still perform replay attacks by using audio commands covertly recoded during a legit interaction of the device owner with the device itself. 

Jang \emph{et al.}~\cite{Jang} suggest the use of a fine-grained access control that gives the user the possibility to grant access to accessibility services for assistive technology apps such as \emph{TextToSpeech} \cite{texttospeech} and \emph{TalkBack} \cite{talkback}. This solution moves the responsibility to the user on deciding if the access is deemed security-sensitive. Unfortunately, a user might not fully understand the security implications of granting such access.

A different approach is proposed by Xu \emph{et al.}~\cite{Xu}, with the design of \emph{SemaDroid}, a privacy-aware sensor management framework for smartphones.  This solution allows users to monitor sensor usage by installed apps and control the disclosure of sensed information. 
 \emph{SemaDroid} can be effective for sensors like GPS and accelerometer that do not require perfect information, but not so for the microphone and speaker. Adjusting the sensing parameters to reduce the quality of sensor data or replacing this data with mock audio data will impair the functionality of audio channels. 
 
Researchers haven also explored improvements to access control mechanisms, but these mechanisms fail to prevent attacks on audio channels.
The main enhancement has been to permit more software to contribute to access control decisions.  
The {\em Android Security Framework} (ASF) \cite{Backes} allows instantiation and deployment of different security models as loadable modules at Android's app layer, middleware and kernel. \system could be deployed as a loadable ASF module, but it would require porting ASF from Android v4.5 to Android v5.0.1. We leave it as future work.
The {\em Android Security Module} (ASM) \cite{Heuser} provides a wide range of security hooks at any level of the Android Stack (Kernel, Middleware, Framework, Services and even Apps).
In ASM, hooks notify the ASM bridge which in turn notifies all the ASM apps that have explicitly registered for those specific hooks.
Therefore, ASM hooks are not suitable for implementing the hooks used by \system since, in \system, apps should not have visibility of the \emph{Audio Hooks} used to mediate access to the microphone and speaker. 


{\em User-Driven Access Control} (UDAC) \cite{Roesner} is the most relevant work for the creation of trusted paths between users and \system. In UDAC, apps use trusted Access Control Gadgets (ACGs) to perform user interaction, which grant apps permission to access user-owned resources or data produced by security-sensitive sensors. Unfortunately, switching the SDK API to require apps to use ACGs would impact millions of apps, not be backward compatible, and lead to false positives when errors in using ACG occur.

\vspace{-0.05in}

\section{Conclusion}
In this paper, we presented \system, a framework that detects and prevents eavesdropping and confused deputy attacks using audio channels.
We identified three different types of audio channels that can be exploited by an adversary: (1) speaker to microphone; (2) speaker to external party; and (3) external party to microphone.
To prevent against attacks, we argue that access control must authorize the information flows resulting from the creation of these audio channels. We then designed \system, an extension to the reference monitor provided by SELinux to enforce lattice policies over dynamically-created audio channels. \system determines the labels associated with all three types of audio channels, enforces information flow policies, and enables resolution of information flow errors.  To enable the resolution of information flow errors, \system uses callbacks to the privileged apps and/or services that are at risk to negotiate acceptable {\em resolvers}, and uses {\em trusted paths} to the user to notify the device owner of risks and gain approval for the creation of audio channels.  
We evaluated \system against six attacks scenarios, covering the three type of audio channels identified, on 17 widely-used apps, showing how \system blocks attacks based on audio channels while preserving functionality for these apps.
\vspace{-0.1in}

\section{Acknowledgments}
Research was sponsored by the Army Research Laboratory and was accomplished under Cooperative Agreement Number W911NF-13-2-0045 (ARL Cyber Security CRA). The views and conclusions contained in this document are those of the authors and should not be interpreted as representing the official policies, either expressed or implied, of the Army Research Laboratory or the U.S. Government. The U.S. Government is authorized to reproduce and distribute reprints for Government purposes notwithstanding any copyright notation here on.
\vspace{-0.1in}


\end{document}